\documentclass{llncs}
\usepackage{makeidx}  
\usepackage{graphicx}
\usepackage{subfigure}
\usepackage{amssymb,amsmath,bm}
\usepackage{mathrsfs}
\usepackage{booktabs}
\usepackage{amsmath}
\usepackage{mathtools}
\usepackage{subfig}
\usepackage[misc]{ifsym}
\usepackage{multirow}
\usepackage{bbding}
\usepackage{array}
 \usepackage{verbatim}
\usepackage{float}
\usepackage{algorithm}
\usepackage{algorithmic}
\usepackage{lettrine}
\usepackage{cite}

\newcolumntype{C}[1]{>{\centering\let\newline\\\arraybackslash}m{#1}}

\usepackage{pbox}

\begin{document}
\frontmatter          
\title{Transformer for Polyp Detection}

\author{\IEEEauthorblockN{AAA\IEEEauthorrefmark{2},
BBB\thanks{* BBB is the corresponding author.}\IEEEauthorrefmark{3}\IEEEauthorrefmark{1}, 
CCC\IEEEauthorrefmark{2}, 
DDD\IEEEauthorrefmark{4}, and
EEE \IEEEauthorrefmark{3}}
\IEEEauthorblockA{\IEEEauthorrefmark{2}xxx, China, \href{mailto:xxx@xxx,xxx@xxx}{\{xxx, xxx\}@xxx}\\
\IEEEauthorrefmark{3}xxx, China, \href{mailto:xxx@xxx,xxx@xxx}{\{xxx, xxx\}@xxx}\\
\IEEEauthorrefmark{4}xxx, China,  \href{mailto:xxx@xxx}{xxx@xxx}}
}

\author{Shijie Liu \and Hongyu Zhou \and Xiaozhou Shi \and Junwen Pan }
\institute{}

\maketitle

\section{Model Architecture}

In recent years, as the Transformer has performed increasingly well on NLP tasks, many researchers have ported the Transformer structure to vision tasks\cite{dosovitskiy2020image, carion2020end, zhu2020deformable, xie2021segformer}, bridging the gap between NLP and CV tasks. In this work, we evaluate some deep learning network for the detection track. Because the ground truth is mask, so we can try both the current detection and segmentation method. We select the DETR\cite{carion2020end} as our baseline through experiment. Besides, we modify the train strategy to fit the dataset.

\begin{figure}[h]
\includegraphics[width=1\textwidth]{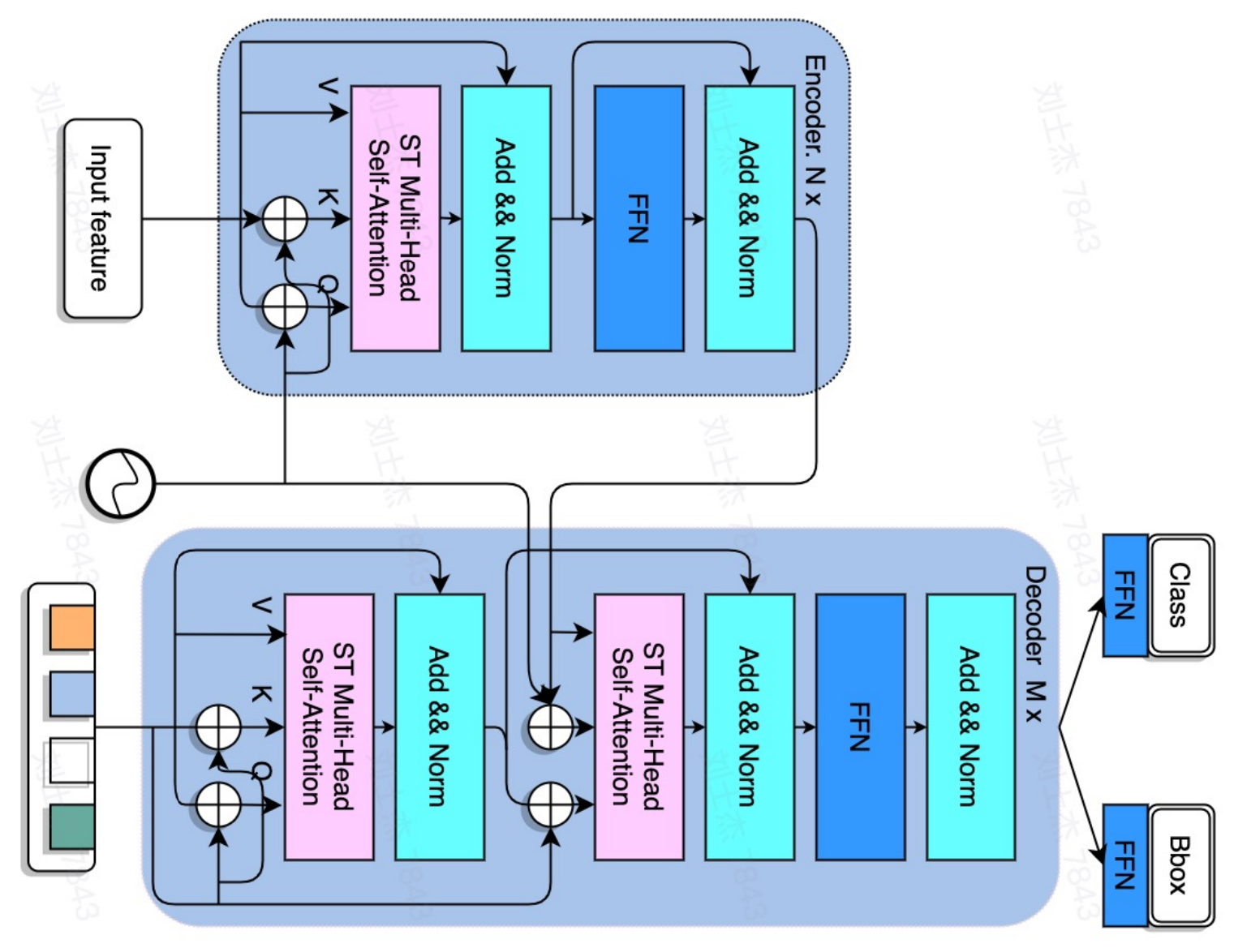}
\caption{Overview Model Architecture}
\label{Fig. 1}
\end{figure}

The overall process is as follows, with DETR (overview fig \ref{Fig. 1}) using a conventional CNN backbone to learn a 2D representation of the input image. Before passing it to the transformer encoder, the model flattens it and complements it with positional encoding. The transformer decoder then takes a small fixed number of learned positional embeddings as input and additionally participates in the processing of the encoder output. We pass each output embedding of the decoder to a shared feedforward network (FFN) that predicts either a detection (class and bounding box) or a "no object" class.

We select the ResNet50 as the backbone to extract the abstract feature. We use the backbone to downsample 32 times, then the features are fed into the downstream encoder. 

In the encoder, the 1x1 convolution reduces the channel dimension of the high-level activation map $f$ from C to a smaller dimension d. A new feature map is created $f_{0} \in R^{d \times H \times W}$. The encoder requires a sequence as input, so we compress the spatial dimensionality of $f_0$ to one dimension, resulting in a $d \times HW $ feature map. Each encoder layer has a standard structure consisting of a multi-headed self-attentive module and a feedforward network (FFN). As the transformer architecture is order-insensitive, we complement it with fixed-position coding, which is added to the input of each attention layer.

The decoder follows the standard architecture of the transformer, using multi-head self-attention and an encoder-decoder attention mechanism to transform N embeddings of size d. The DETR is different with the standard orginal transformer, which decodes N objects in parallel at each decoder level. For each detr decoder there are two inputs, one is the output from the encoder and the other is the Object query to be learned. Besides, the decoder of the detr add the spatial positional encoding. In the classifier section, the output embedding transformed by the decoder will feed into the FFN to decode the bounding box coordinate and class labels.

The FFN is combined by 4 fc layers with ReLU activation function, which decode the output of last transformer decoder separately. The regression layer output the scaled object center coordinate and the object's height and width, and the classify layer predict the object's label using softmax functions. The detr designed the fixed number of bounding box for every image, network will predict 100 bounding box and score. 

\section{Model Training}
\subsection{Augumentation and Fine-Tuning}
Our model is based the PyTorch toobox. In order to train our model faster and with better generalisation performance, we fine-tune our algorithm on a pre-trained network of the coco dataset. During the training process, we found that bubbles or highlights in the video would cause the model to misidentify them as polyps. For this case, we used a number of data enhancements to simulate these noises, including Random Brightness, ColorJitter, GaussianBlur, RandomFlip, Random Sharpness and RandomResizedCrop. Besides, we have also blurred some of the images in the training set so that the model can recognise them even for blurred and confusing situations. Because polyps vary greatly in size and have diverse morphological variations. Therefore, we chose a multi-scale training approach to adapt the model to the diversity of the learned lesions. We set up a strategy of multi-scales input and randomly selecting the size of the output to achieve this. We used scaling enhancement to resize the input image so that the short edge was at least 288 pixels and at most 416 pixels, and the long edge was at most 512. For the training process we selected the Adam optimizer and set the initial learning rate to 0.0001 and weightdecay to 0.001. We select the box with a score greater than 0.3 as the final output.

\subsection{Loss function}
DETR model will predict a set of N(follow the default 100) predictions. We can obtain the FFN's prediction including class and bbox. The next step is to calculate the cost of the Hungarian match, the goal is to minimise margin errors while ensuring that the categories are correct. This process can format as following:

\begin{equation}
\mathcal {\hat{\sigma}} = argmin \sum_{i}^{N} L_{match}(y_i, \hat{y}_{\sigma(i)},
\end{equation}
\begin{equation}
\mathcal L_{match} = -1_{c_i \ne \emptyset} {\hat{p}}_{\sigma(i)}(c_i) + 1_{c_i \ne \emptyset}L_{box}(b_i, \hat{b}_{\sigma(i)}),
\end{equation}

where $y_i$ is gt and $\hat(y)_\sigma(i)$ is prediction. For the prediction with index $\sigma_i$ we define probability of class $c_i$ as $\hat{p}_{\sigma(i)}(c_i)$ and the predicted box as $\hat{b}_{\sigma(i)}$.

Taking into account the distance between the category and bbox, and adding a cross-comparison ratio using giou to the cost of the Hungarian match. The last can format as following:  

\begin{equation}
\mathcal L_{Hungarian}(y, \hat{y}) = \sum_{i=1}^N -log {\hat{p_{\sigma(i)}(c_i)}} + 1_{c_i \ne \emptyset}L_{box}(b_i, \hat{b}_{\hat{\sigma}(i)}),
\end{equation}

\begin{equation}
\mathcal L_{box}(b_i, b_{\sigma(i)}) = \lambda_{iou}(b_i, \hat{b}_{\sigma(i)}) + \lambda_{L1}||b_{i} - b_{\sigma(i)}||
\end{equation}

\section{Example of Results}
\begin{figure}[h]
\includegraphics[width=1\textwidth]{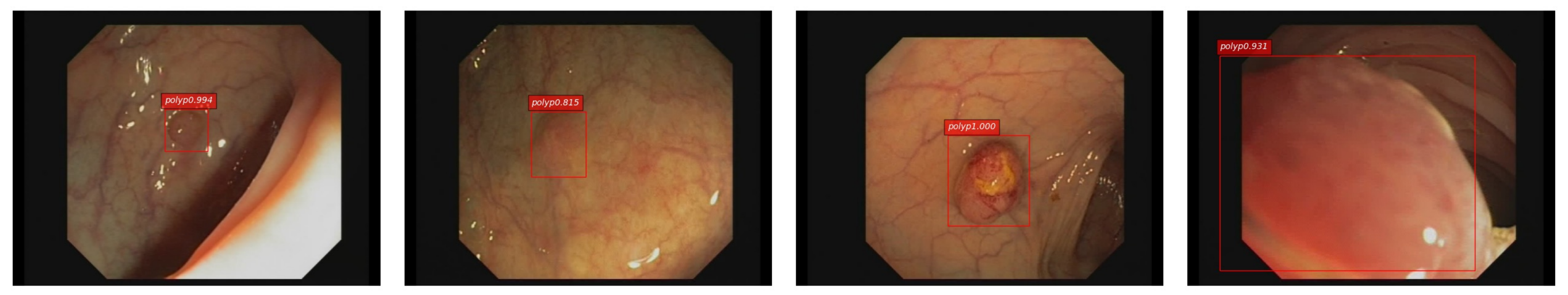}
\caption{Test Result Visualization(small/media/large)}
\label{Fig. 2}
\end{figure}
	
To test the validity of the model and training schedule, we used a five-fold cross-validation on the training-val set. The following Fig. \ref{Fig. 2} shows the visualization results of some representative samples in the test set.

\bibliographystyle{plain}
\bibliography{age}

\end{document}